\documentclass[preprint, 11pt]{JHEP3} 

\JHEPspecialurl{http://jhep.sissa.it/JOURNAL/JHEP3.tar.gz}

\usepackage{epsfig,multicol,bbm}
\usepackage{epstopdf}

\newcommand\fverb{\setbox\fverbbox=\hbox\bgroup\verb}
\newcommand\fverbdo{\egroup\medskip\noindent%
			\fbox{\unhbox\fverbbox}\ }
\newcommand\fverbit{\egroup\item[\fbox{\unhbox\fverbbox}]}
\newbox\fverbbox

\def\fig#1{{Fig.~\ref{#1}}}
\def\app#1{{App.~\ref{#1}}}

\title{Momentum tensors of second and third rank, as tools for jets' analysis}

\author{Samuel Dagan\thanks{Email
: dagan@post.tau.ac.il}\hspace{.1in}and Arie Vainshtein\thanks{Email: varie888@gmail.com} \\ School of Physics and Astronomy, Tel-Aviv University, Tel-Aviv 69978, Israel} 

\received{\today} 		

\preprint{TAUP - 2927/10}

\abstract{Rank two momentum tensors, free of mass singularities, are often used for identification and analysis of two and three hadronic jet events. Based on Monte-Carlo generated events, we find that for sufficiently high energies, the tensor invariants provide the characteristics of a three-jet event, without assigning the particles to the jets. Furthermore, the combined application of rank two and rank three momentum tensors, provides a signature for a 3-jet event.}


\begin{document} 

\section{Introduction}\label{intro}

The run of the Large Hadronic Collider at CERN with energies, at least an order of magnitude higher than the previous colliders, poses new challenges to the high energy physics community. One of them, faced by the experimentalists, is the handling of very high multiplicity events. The time needed to analyze such events grows more than linearly, when fitting procedures are employed, for instance - jet  identification by clustering.

On the other hand, as it will be shown, the higher the energies of the jets are, the measured accuracy of the momentum tensors' global variables increases, making them more suitable for analysis.

Two quark-antiquark ($q\bar{q}$) jets were first seen in $e^{+}e^{-}$ events \cite{Hanson}, as two clusters of particles moving back to back in opposite directions, with the same absolute value of momentum.
\FIGURE{\epsfig{file=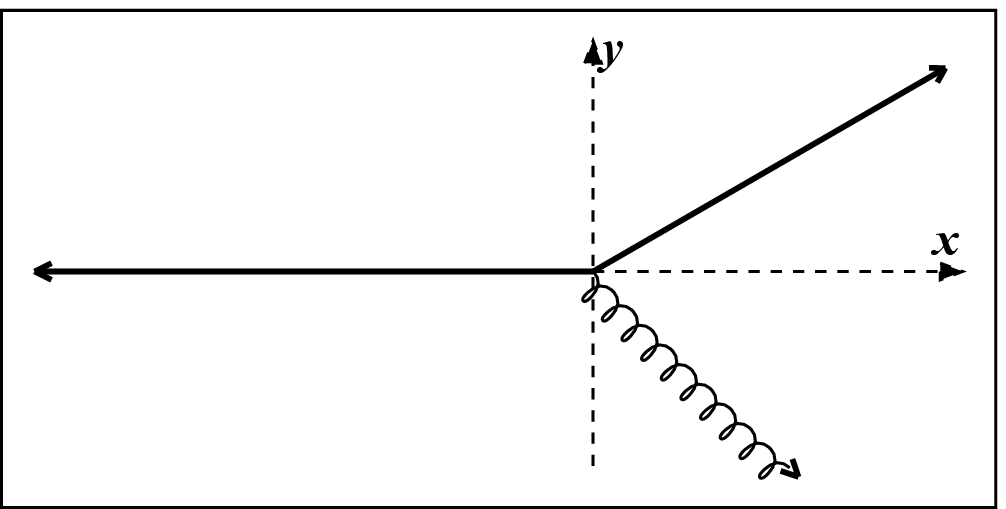,width=70mm}
\caption{Schematic view of a 3-jet event}
\label{3jets1}
}
Three jets, with a gluon jet radiated from one of the quark jets, were observed later on, also in $e^{+} e^{-}$ events \cite{Barber,Berger,Brandelik}. This is illustrated in \fig{3jets1}. Similarly to two jets, the three jets appear in their centre of mass system with vanishing total momentum.

For the study of jet events, a symmetric tensor of rank two, formed by the momenta of the particles, was suggested by~\cite{Bjorken}. It was studied and modified later on by~\cite{Parisi} and~\cite{Donoghue}, in order to obtain a tensor, free of mass singularities, and therefore computable by perturbative QCD:
\begin{equation}
Q_{ij}  = {{\sum\limits_{n = 1}^N {{{p_{n,i} p_{n,j} } \over {\left| {p_n } \right|}}} } \over {\sum\limits_{n = 1}^N {\left| {p_n } \right|} }} \label{1.01}
\end{equation}
The summation over {\emph n} is up-to the total number of particles {\emph N},  $p_n$  is the momentum of the particle {\emph n} and the indices {\emph i} and {\emph j} indicate components of the momenta. The denominator serves for normalization.

The reference \cite{Donoghue} actually gives a general definition for such a symmetric momentum tensor of any rank {\emph r}:
\begin{equation}
Q_{ijk...r}  = {{\sum\limits_{n = 1}^N {{{p_{n,i} p_{n,j} p_{n,k} ...p_{n,r} } \over {\left| {p_n } \right|^{r - 1} }}} } \over {\sum\limits_{n = 1}^N {\left| {p_n } \right|} }} \label{1.03}
\end{equation}
Note that the normalization factor is identical to that of rank 2, and the terms of the tensor are linear with respect to momenta.

In addition to the important theoretical features, it can be easily shown that for any rank, the tensor remains invariant if one particle is considered as two collinear ones, or if two collinear particles appear as - one. Such errors occur frequently, when calorimeter data is used for neutral particles.

Using the common convention that a repeated index means summation, one obtains that the trace of the tensor of rank two~(\ref{1.01}) is:
\begin{equation}
Q_{ii}  =  \lambda_1+\lambda_2+\lambda _3 = 1 \label{1.04}
\end{equation}
where the $\lambda$'s are the eigenvalues. By writing them in decreasing order:
\begin{equation}
\lambda _1  \ge \lambda _2  \ge \lambda _3 \label{1.05}
\end{equation}
and by defining an \emph {idealized jet} with collinear particles only, meaning that a jet is expressed by a single momentum, one obtains that an event with two such jets has
\begin{equation}
\lambda _1  = 1\quad \quad \lambda _2  = \lambda _3  = 0 \label{1.06}
\end{equation}
and any idealized three jet event is planar:
\begin{equation}\lambda _3  = 0\quad \quad \lambda _1  \ge \lambda _2  > 0\quad \quad \lambda _1  + \lambda _2  = 1 \label{1.07}
\end{equation}
Note though, that not all planar events correspond to three jets.

The remnants of the beam particles \emph{in hadron-hadron collisions} carry part of the energy and momentum, before disappearing inside the collision pipe. As a result, the two-jet events are not back to back, and the three jet events are not planar anymore. In order to use the same tensors in the analysis, \emph{it is imperative to make the appropriate Lorentz boost before-hand, by requiring that the total momentum of the particles, originating from the event's vertex, vanishes as a result of the transformation}. 

For a more detailed study of three and more jet events, clustering algorithms were developed. A review article of algorithms for  $e^{+}e^{-}$ events can be seen in~\cite{Moretti}. Algorithms for hadron-hadron events have been more recently developed, e.g~\cite{Cacciary}. A fitting procedure is used, which could be time consuming for high multiplicity events, as those obtained at the LHC. 
The present study could facilitate this procedure in the case of 3-jet events, by doing a priory selection. In addition it could yield some independent measurements for comparison.

This paper is organized as follows. {\bf Section \ref{ideal}} deals with idealized jets and their properties, analyzed by the momentum tensors of ranks two and three. {\bf Section \ref{real}} deals with more realistic jets, generated by Monte-Carlo, and concludes with suggested implementations for real data. {\bf Section \ref{summary}} contains a summary and conclusions.

\section{Idealized jets}\label{ideal}

An \emph {idealized jet}, as defined in the introduction, has only collinear particles and is expressed by a single momentum. 

\subsection{Tensor of rank two}\label{ideal2t}
In the case of a planar event, the eigenvalues of the rank-2 tensor of momenta~(\ref{1.01}) fulfill the relations~ (\ref{1.07}). Therefore, only one free parameter defines the global event shape, which is chosen here as: 
\begin{equation}
\mu  = \lambda _1 \lambda _2 \quad \quad 0 < \mu  \le {1 \over 4} \label{2.01}
\end{equation}
Details of the physical meaning of the tensor's invariants can be found in \cite{Ellis}. 
\FIGURE{\epsfig{file=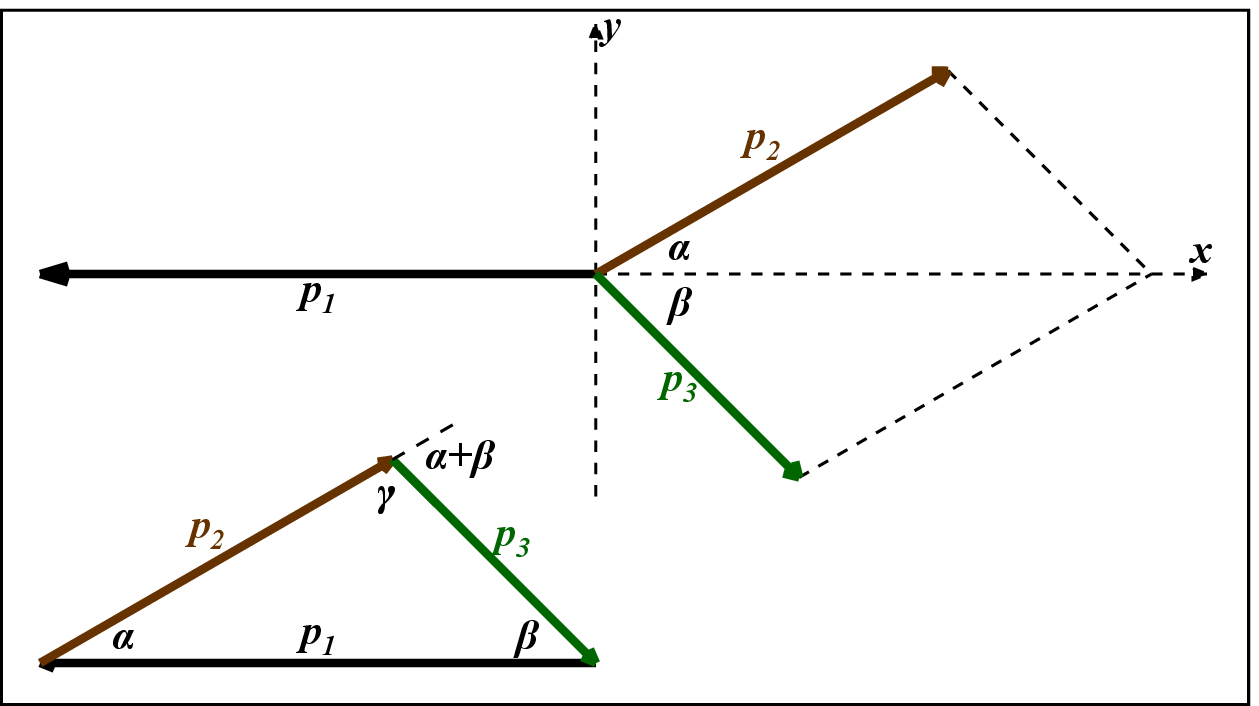,width=115mm}
\caption{Idealized 3-jet event.}
\label{3jets2}
}

The simulation of an idealized 3-jet event is done here in the ({\emph x,y}) plane by the momenta components of the jets:
\begin{equation}
{|{p_1 \left( { - p_1 ,0} \right)}|} \ge {|{p_2 \left( {p_2 \cos \alpha ,p_2 \sin \alpha } \right)}|} \ge {|{p_3 \left( {p_3 \cos \beta , - p_3 \sin \beta } \right)}|} > 0 \label{2.02}
\end{equation}
where the size of momenta are ordered for convenience.
From the triangle of the momenta in \fig{3jets2} one obtains:
\begin{equation}
{{| {p_1 }|} \over {\sin \left( {\alpha  + \beta } \right)}} = {{| {p_2 } |} \over {\sin \beta }} = {{| {p_3 } |} \over {\sin \alpha }} = p_0 \label{2.03} 
\end{equation}
where ${\emph p}_0$ is a free parameter of no importance and the angles fulfill:
\begin{equation}
0 < \alpha  \leq \beta  \leq \gamma  = \pi  - \alpha  - \beta \label{2.04} 
\end{equation}

The none-vanishing terms of the momentum tensor~(\ref{1.01}) corresponding to~(\ref{2.02}) are:
\begin{equation}
\left. \matrix{
  DQ_{11}  = D - \sin \alpha \sin \beta \left( {\sin \alpha  + \sin \beta } \right)\; \hfill \cr 
  DQ_{22}  = \sin \alpha \sin \beta \left( {\sin \alpha  + \sin \beta } \right) \hfill \cr 
  DQ_{12}  = \sin \alpha \sin \beta \left( {\cos \alpha  - \cos \beta } \right) \hfill \cr 
  D = \sin \alpha  + \sin \beta  + \sin \left( {\alpha  + \beta } \right) \hfill \cr}  \right\}
\label{2.05}
\end{equation}
After some trigonometric manipulations one obtains:
\begin{equation}
\mu  = \lambda _1 \lambda _2  = \left| {\matrix{
   {Q_{11} } & {Q_{12} }  \cr 
   {Q_{21} } & {Q_{22} }  \cr 
 } } \right| = {{\sin \alpha \sin \beta \sin \left( {\alpha  + \beta } \right)} \over {\sin \alpha  + \sin \beta  + \sin \left( {\alpha  + \beta } \right)}}
\label{2.06}
\end{equation}
with eigenvalues of the tensor:
\begin{equation}
\lambda _{1,2}  = {{1 \pm \sqrt {1 - 4\mu } } \over 2} \label{2.07}
\end{equation}
where $\mu$ from~(\ref{2.06}) can be rewritten also as:
\begin{equation}
\mu  = \cos {\sigma  \over 2}\left( {\cos {\delta  \over 2} - \cos {\sigma  \over 2}} \right)\quad \quad {\rm{where}}\quad \left\{ \matrix{
  \sigma  = \beta  + \alpha  \hfill \cr 
  \delta  = \beta  - \alpha  \hfill \cr}  \right\} \label{2.08}
\end{equation}
\FIGURE{\epsfig{file=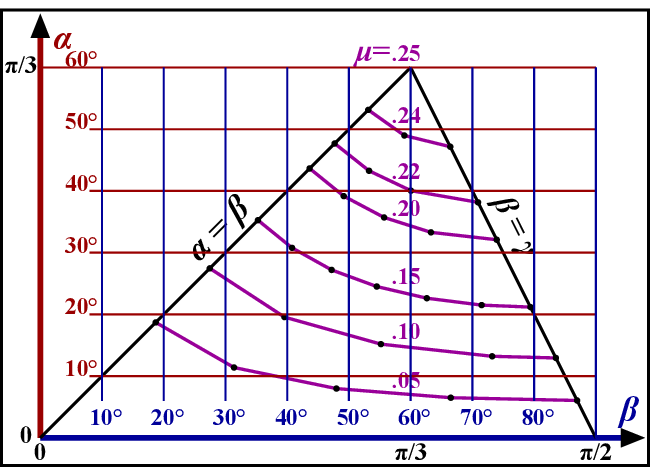,width=110mm}
\caption{Level lines of $\mu$. See (\ref{2.01}), (\ref{2.02}) and (\ref{2.04}) for details.}
\label{3jets3}}
The domain of the angles $\alpha$ and $\beta$ is displayed in  \fig{3jets3} together with some of the curves corresponding to constant $\mu$ values. The dots represent calculated values connected by straight lines. The following relations correspond to the boundaries:
\begin{equation}
\left. \matrix{
  \alpha  = \beta \quad \quad  \Rightarrow \quad \quad \cos \beta  = {{1 + \sqrt {1 - 4\mu } } \over 2} = \lambda _1 \quad \quad \& \quad \quad {|{p_2}|}  = {|{p_3}|}  \hfill \cr 
  \beta  = \gamma \quad \quad  \Rightarrow \quad \quad \cos \beta  = {{1 - \sqrt {1 - 4\mu } } \over 2} = \lambda _2 \quad \quad \& \quad \quad {|{p_1}|}  = {|{p_2}|}  \hfill \cr}  \right\} \label{2.09}
\end{equation}
The value of $\mu$, does not allow an unambiguous determination of the angles, except in the case of $\mu=0.25$. The value of $\mu=0$ corresponds to a two-jet event, and therefore very small $\mu$ values are not a good indication for a three-jet event. For a given $\mu$ value, the angles $\alpha$ and $\beta$ are constrained by the relation (\ref{2.08}), and bounded to the values of (\ref{2.09}). 

The $\mu$ values of planar events are bounded by~(\ref{2.01}), exactly as in the case of three ideal jets. In the latter case the maximal $\mu$ value corresponds to rotationally symmetric distribution of the jets. Let's define an idealized, rotationally symmetric planar event of $N$ jets, as having identical absolute values of momenta, and each jet being separated by an angle of $2\pi/N$ from its neighbours. 
\FIGURE{\epsfig{file=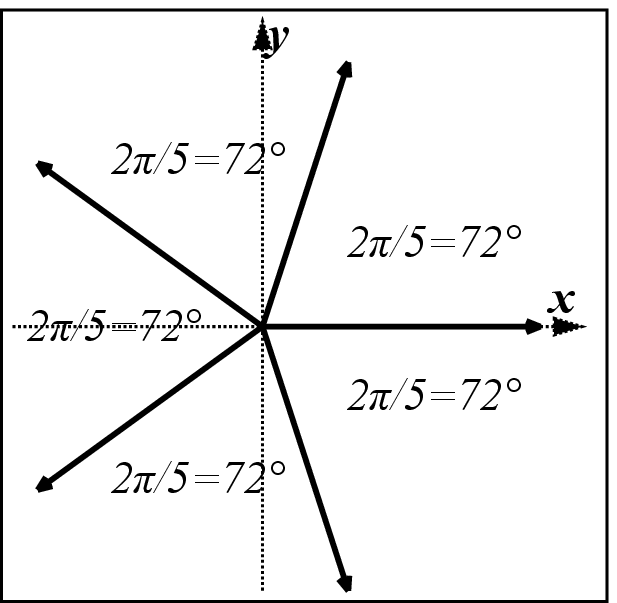,width=65mm}
\caption{Symmetric planar event of five jets}
\label{5jets}}
An example of $N=5$ is illustrated in \fig{5jets}, and the momenta of the jets for any $N\geq3$ are expressed by
\begin{equation}
\left. \matrix{
  \overrightarrow {p_n }  = \left[ {p_0 \cos \left( {{{2\pi n} \over N}} \right)\,,\,p_0 \sin \left( {{{2\pi n} \over N}} \right)} \right] \hfill \cr 
  n = 1\,,\,...\,,N\quad \quad N \ge 3 \hfill \cr}  \right\} \label{symNjets}
\end{equation}
In such a case, because of the symmetry, the eigenvalues are equal ($\lambda _1 = \lambda _2$), and therefore from (\ref{2.01})   and (\ref{1.07}), $\mu$ becomes maximal:
\begin{equation}
\mu  = {1 \over 4}
\label{muMax}
\end{equation}
This result can be obtained of course also by direct calculation, which is shown in \app{signature2}, in order to be compared later with tensor of rank three.

\subsection{Tensor of rank three} \label{ideal3t}
From the definition of linear tensors~(\ref{1.03}), the tensor of rank three is:
\begin{equation}
Q_{ijk}  = {{\sum\limits_{n = 1}^N {{{p_{n,i} p_{n,j} p_{n,k} } \over {\left| {p_n } \right|^2 }}} } \over {\sum\limits_{n = 1}^N {\left| {p_n } \right|} }} \label{3.01}
\end{equation}
and is used in \cite{Donoghue} as an example of QCD calculation for 3-jet events. As far as we know, it has not yet been used as a tool for experimental analysis.

Contrary to the tensor of rank two, and due to the products of odd number of momenta, this tensor vanishes for any idealized two-jet event. This property, together with the tensor of rank two, can be used to enhance the signature of a two-jet event.

From~(\ref{3.01}) and using the summation convention one obtains a vector:
\begin{equation}
V_j  = Q_{jii}  = {{\sum\limits_{n = 1}^N {{{p_{n,j} \left( {p_{n,x}^2  + p_{n,y}^2  + p_{n,z}^2 } \right)} \over {\left| {p_n } \right|^2 }}} } \over {\sum\limits_{n = 1}^N {\left| {p_n } \right|} }} = {{\sum\limits_{n = 1}^N {p_{n,j} } } \over {\sum\limits_{n = 1}^N {\left| {p_n } \right|} }} = 0 \label{3.02}
\end{equation}
that vanishes, since it is the total momentum of the particles in the centre of mass system. This yields three linear relations between the terms of the tensor:
\begin{equation}
\left. \matrix{
  Q_{111}  + Q_{122}  + Q_{133}  = 0\; \hfill \cr 
  Q_{211}  + Q_{222}  + Q_{233}  = 0\; \hfill \cr 
  Q_{311}  + Q_{322}  + Q_{333}  = 0\; \hfill \cr}  \right\} \label{3.03}
\end{equation}
and of course the absolute value of the vector also vanishes:
\begin{equation}
V_j V_j  = 0 \label{3.04}
\end{equation}

One can also obtain a tensor of rank-2:
\begin{equation}
R_{ij}  = Q_{imn} Q_{jmn} \label{3.05} 
\end{equation}
which yields:
\begin{equation}
\left. \matrix{
  R_{11}  = Q_{111}^2  + Q_{122}^2  + Q_{133}^2  + 2\left( {Q_{211}^2  + Q_{123}^2  + Q_{311}^2 } \right) \hfill \cr 
  R_{22}  = Q_{222}^2  + Q_{211}^2  + Q_{233}^2  + 2\left( {Q_{122}^2  + Q_{123}^2  + Q_{322}^2 } \right) \hfill \cr 
  R_{33}  = Q_{333}^2  + Q_{311}^2  + Q_{322}^2  + 2\left( {Q_{133}^2  + Q_{123}^2  + Q_{233}^2 } \right) \hfill \cr 
  R_{12}  = Q_{133} Q_{233}  - Q_{133} Q_{211}  - Q_{122} Q_{233}  - 2Q_{123} Q_{333}  \hfill \cr 
  R_{23}  = Q_{211} Q_{311}  - Q_{211} Q_{322}  - Q_{233} Q_{311}  - 2Q_{123} Q_{111}  \hfill \cr 
  R_{31}  = Q_{322} Q_{122}  - Q_{322} Q_{133}  - Q_{311} Q_{122}  - 2Q_{123} Q_{222}  \hfill \cr}  \right\} \label{3.06} 
\end{equation}

For planar events, assuming e.g.  the $(x,y)$ plane, any term of the $Q_{ijk}$ tensor vanishes, if at least one of its indices equals 3. From~(\ref{3.06}) one obtains in such a case that there are only two non-vanishing terms of $R_{ij}$:
\begin{equation}
R_{11}  = Q_{111}^2  + Q_{122}^2  + 2Q_{211}^2  = 2\left( {Q_{122}^2  + Q_{211}^2 } \right) = 2\left( {Q_{111}^2  + Q_{222}^2 } \right) = R_{22} \label{3.07}  
\end{equation}
where the relations of (\ref{3.03}) were used. If the eigenvalue of the tensor $R_{ij}$ are denoted by the letter $\nu$ and are ordered by decreasing values, in case of planar event one obtains:
\begin{equation}
\nu _1  = \nu _2  \ge \nu _3  = 0 \label{3.08}  
\end{equation}
\emph{For a simulated idealized 3-jet event in the $(x,y)$ plane, one obtains} from (\ref{3.07}) and (\ref{3.08}):
\begin{equation}
Q_{111}^2  + Q_{222}^2  = \mu ^2 
\label{3.09}   
\end{equation}
\emph{where $\mu$ has exactly the same value as in the case of (\ref{2.06})}. Using the eigenvalues one can summarize that \emph{for an idealized 3-jet event}:
\begin{equation}
\left. \matrix{
  \nu _1  = \nu _2  = 2\lambda _1^2 \lambda _2^2  = 2\mu ^2  \hfill \cr 
  \nu _3  = \lambda _3  = 0 \hfill \cr}  \right\} \label{3.10}
\end{equation}

This is a \emph{genuine signature of a three-jet  event, which was obtained by two entirely different mathematical mechanisms}, while in general (\ref{3.09}) does not hold. As an example, any planar rotationally symmetric event of $N\geq3$ jets fulfills $\lambda _1 \lambda _2  = \mu  = 0.25$ (\app{signature2}), but only for $N=3$, (\ref{3.10}) takes place, while for $N>3$:  $\nu _1  = \nu _2  = 0$ (\app{signature3}).

\section{Real jets} \label{real}

In a real event, due to the hadronization process, the momenta of the particles of a jet are close to the jet axis direction, but are not necessarily collinear with it. Consequently the planar property of a 3-jet event is broken, i.e. the smallest eigenvalues of momentum tensors, $\lambda_3$ and $\nu_3$ (\ref{3.10}), do not vanish any-more. 

In order to study this behaviour, 3-jet Monte-Carlo events were generated and particles produced by the hadronization process. \emph{The momenta and energies of the primary particles obtained from the hadronization were recorded in their common centre of mass system}. They were used for the calculation of the momentum tensors, and for comparison with the idealized case. \emph{By definition, this system is equivalent to the common centre of mass of the jets}. 

We have used version 6.408  \cite{Pythia} of the  PYTHIA program to generate high-energy physics events. The given total energy, in the centre of mass system of the jets, was randomly distributed among the three jets. The most energetic jet was assigned to a quark jet, while the other two - to a quark and to a gluon jet randomly. 
The quark jets originated from  $u$ or $d$ quarks. Events with a jet energy less than 2 GeV where not accepted, which resulted in the suppression of events with $\mu \le$ 0.05, less pronounced at higher c.m. energies. 

Some of the parameters, corresponding to the idealized events, can be obtained with a good approximation directly from the real events. Other require fitting and parameterization depending on the choice of the event generation mechanism, which is very simple here. For any other choice, the same procedure can be easily repeated, and the appropriate parameters obtained.

\subsection{Tensor of rank two}
It is natural to start by asking how to obtain a measured value of $\mu_1$, close enough to the idealized value of $\mu$~(\ref{2.01}). Let's consider two definitions of $\mu_1$:  $\mu'_1$ and $\mu''_1$ (\ref{4.10}). Since $\lambda_3\neq0$ for real events, $\mu'_1$ is smaller than  the idealized $\mu$, due to the trace invariance (\ref{1.04}). 
\begin{equation}
\mu'_1 = \lambda _1 \lambda _2 \quad \leq \quad  \mu \quad \leq \quad \lambda _1 \lambda _2  + \lambda _2 \lambda _3  + \lambda _3 \lambda _1  = {C \over 3} = \mu''_1 
\label{4.10}
\end{equation}
The $\mu''_1$ is actually a tensor invariant, well studied in \cite{Ellis} as the C parameter. For the real case, this value is bigger than that of the idealized case, due to the hadronization. For the idealized case, it  also corresponds to $\mu$~(\ref{2.01}).
\FIGURE{\epsfig{file=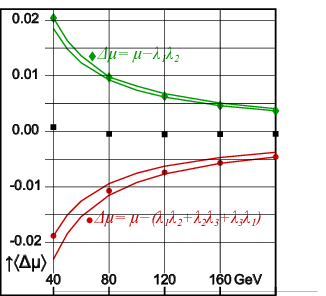,width=70mm}
\caption{Distribution of the mean values of $\Delta\mu$ (ordinate), as defined in the figure,  vs the c.m. energy of three jets (abscissa). The square-shaped signs indicate the average $\Delta\mu$ of both distributions. The curves are explained in the text.}
\label{dele2}}
The distributions of the mean $\Delta \mu  = \mu  - \mu_1$ values, where $\mu$ corresponds to the idealized 3-jet events, and $\mu_1$  to $\mu'_1$ or to $\mu''_1$ (\ref{4.10}), obtained from generated 40000 events for each one of the five c.m. energy $E$, are shown in \fig{dele2}. Two different signs mark the data points.  For $\mu''_1$ (bottom), the values of  $\Delta \mu$ are called hadronization corrections. The hadronization corrections of the event shape parameters, e.g. C (\ref{4.10}), have been studied first by \cite{Webber} and \cite{Dokshitzer+} and developed further by more authors. Reference~\cite{Dokshitzer} contains a review of these studies . The bottom distribution of \fig{dele2} should behave approximately as
\begin{equation}
\Delta \mu  = \mu  - \left( {\lambda _1 \lambda _2  + \lambda _2 \lambda _3  + \lambda _3 \lambda_1 } \right) =  - {a \over E}  \label{4.101}
\end{equation}
where $a$ is a positive constant. The two solid curved lines express the boundaries of the data points, where the extreme values of $a$ are used:
\begin{equation}
0.75 \le a \le 0.92\;{\rm{GeV}} \label{4.102}
\end{equation}
The top distribution of \fig{dele2} behaves similarly, but with inverse sign, yielding:
\begin{equation}
\left. \matrix{
  \Delta \mu  = \mu  - \lambda _1 \lambda _2  = {b \over E} \hfill \cr 
  0.74 \le b \le 0.82\;{\rm{GeV}} \hfill \cr}  \right\} \label{4.103}
\end{equation}
The values of $a$ (\ref{4.102}) and $b$ (\ref{4.103}) are close to each other, and the graphical displays behave almost as mirror images. 

This behaviour can be explained by simple considerations. An idealized ($i$) 3-jet event in the ($x,y$) plane fulfills (\ref{lamideal}). \begin{equation}\lambda _3(i)  = 0\quad \quad \lambda _1(i)  \ge \lambda _2(i)  > 0\quad \quad \lambda _1(i)  + \lambda _2(i)  = 1 \label{lamideal}
\end{equation}
After hadronization, vector components outside the plane appear, and one obtains:
\begin{equation}
\lambda _{3}  = \Delta _3  > 0\quad  \Rightarrow \quad \lambda _{1}  = \lambda _1(i)  - \Delta _1 \quad {\rm{and}}\quad \lambda _{2}  = \lambda _2(i)  - \Delta _2 = 1 - \lambda_1(i) - \Delta _2 \label{hadron}
\end{equation}
From the trace invariance (\ref{1.04}), one obtains:
\begin{equation}
\Delta _3  = \Delta _1  + \Delta _2 \label{Del3}
\end{equation}
For given values of $\mu$ and $E$, the values of $\Delta_1$ and $\Delta_2$ could in principle differ one from the other, but by
symmetry considerations we assume that \emph{their means} are equal: 
\begin{equation}
\Delta _1  = \Delta _2  = \Delta \quad {\rm{and}}\quad \Delta _{\rm{3}}  = 2\Delta 
 \label{DelAve}
\end{equation}
Assuming that $\Delta$ is much smaller than one, we can write that
\begin{equation}
\Delta ^2  \ll \Delta  
 \label{DelSquare}
\end{equation}
yielding the first approximations of the mean values
\begin{equation}
\left. \matrix{
  \mu '_1  = \lambda _1 \lambda _2  = \left[ {\lambda _1 \left( i \right) - \Delta } \right]\left[ {\lambda _2 \left( i \right) - \Delta } \right] = \mu  - \Delta \left[ {\lambda _1 \left( i \right) + \lambda _2 \left( i \right)} \right] + \Delta ^2  = \mu  - \Delta  \hfill \cr 
  \mu ''_1  = \mu '_1  + \lambda _3 \left( {\lambda _1  + \lambda _2 } \right) = \mu '_1  + \lambda _3 \left( {1 - \lambda _3 } \right) = \mu  - \Delta  + 2\Delta \left( {1 - 2\Delta } \right) = \mu  + \Delta  \hfill \cr}  \right\}
\label{mu'andmu''}
\end{equation}
The $\Delta$ from (\ref{mu'andmu''}) is the mean value of the hadronization correction for 3-jet events, which can be measured experimentally by
\begin{equation}
\Delta  = {{\mu ''_1  - \mu '_1 } \over 2} = {{\lambda _3 \left( {\lambda _1  + \lambda _2 } \right)} \over 2} = {{\lambda _3 \left( {1 - \lambda _3 } \right)} \over 2}
\label{apprhadcorr}
\end{equation}
An approximation of the idealized $\mu$ value can be obtained experimentally by the mean value of $\mu_1$:
\begin{equation}
\mu_1  = {{\mu ''_1  + \mu '_1 } \over 2} = {{2\lambda _1 \lambda _2  + \lambda _3 \left( {\lambda _1  + \lambda _2 } \right)} \over 2} = {{2\lambda _1 \lambda _2  + \lambda _3 \left( {1 - \lambda _3 } \right)} \over 2}
 \label{apprmu}
\end{equation}
\DOUBLEFIGURE[ht]{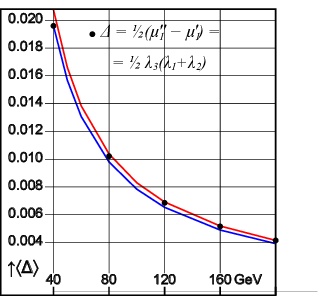,width=70mm}
{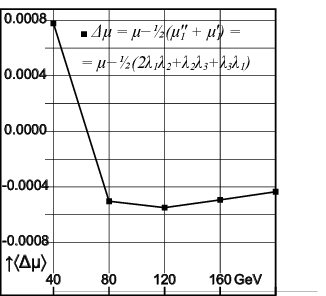,width=70mm}
{Distribution of the mean values of  the hadronization corrections $\Delta$ (ordinate), defined in the figure,  vs the c.m. energy of three jets (abscissa). The curves represent the boundaries of the points. \label{hadcor}}
{Distribution of the mean value of $\Delta\mu$ (ordinate), defined in the figure, vs the c.m. energy (abscissa) of three jets. The straight lines connecting the data points are just for guiding the eye. \label{dele1}}

The hadronization corrections $\Delta$ (\ref{apprhadcorr}), averaged over different energies , are plotted in \fig{hadcor}. The curves that bound the plotted points are 
\begin{equation}
{{0.78} \over E} \le \Delta  = {\alpha  \over E} \le {{0.83} \over E}
 \label{hadcorlimit}
\end{equation}
in agreement with (\ref{4.101}) - (\ref{4.103}). Here they represent the values obtained from the event generator used. In the case of experimental data of 3-jet events, they should indicate the measured value of the hadronization corrections for $\mu={C \over 3}$  (\ref{4.10}) .

\FIGURE{\epsfig{file=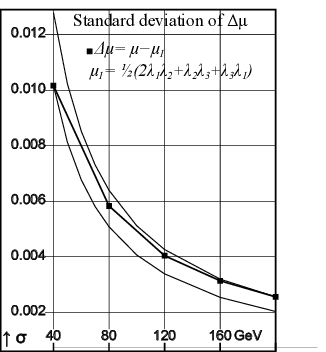,width=70mm}
\caption{Distribution of the standard deviation $\sigma$ of $\Delta\mu$ (ordinate) defined in the figure vs the c.m. energy (abscissa). The straight lines connecting the data points are just for guiding the eye.} 
\label{sige1}}

The  arithmetic average of the upper and bottom distributions of \fig{dele2}, displayed by squared marks, corresponds to the $\Delta\mu=\mu  - \mu _1$ (\ref{apprmu}) distributions. The enlarged-scale distribution of the latter is given in \fig{dele1}.

From the same sample of events, the standard deviations of the distributions were calculated and plotted in \fig{sige1}. They exhibit values, more than one order of magnitude larger than the corresponding $\Delta\mu$ values of \fig{dele1}, therefore the choice of 
(\ref{apprmu}) is acceptable as an approximation to the idealized $\mu$ value. 
The solid curves from \fig{sige1} bounding the distribution of the standard deviation are 
\begin{equation}
\left. \matrix{
\sigma \left( {\Delta \mu } \right) = {d \over E} \hfill \cr 
 0.40 \le d \le 0.51\;{\rm{GeV}} \hfill \cr}  \right\} \label{4.112}
\end{equation}

 \fig{mu80} and \fig{mu200} show scatter plots of the values of $\mu$ (ordinate) vs $\mu_1$ (\ref{apprmu}) (abscissa) obtained from tree-jet generated events, for two c.m. energies: 80 and 200 GeV. Note the depletion of events with at $\mu<0.05$, mentioned \textref{real}{previously},  and the narrowing of the distribution with energy and with the increase of the $\mu$ values .
\DOUBLEFIGURE[ht]{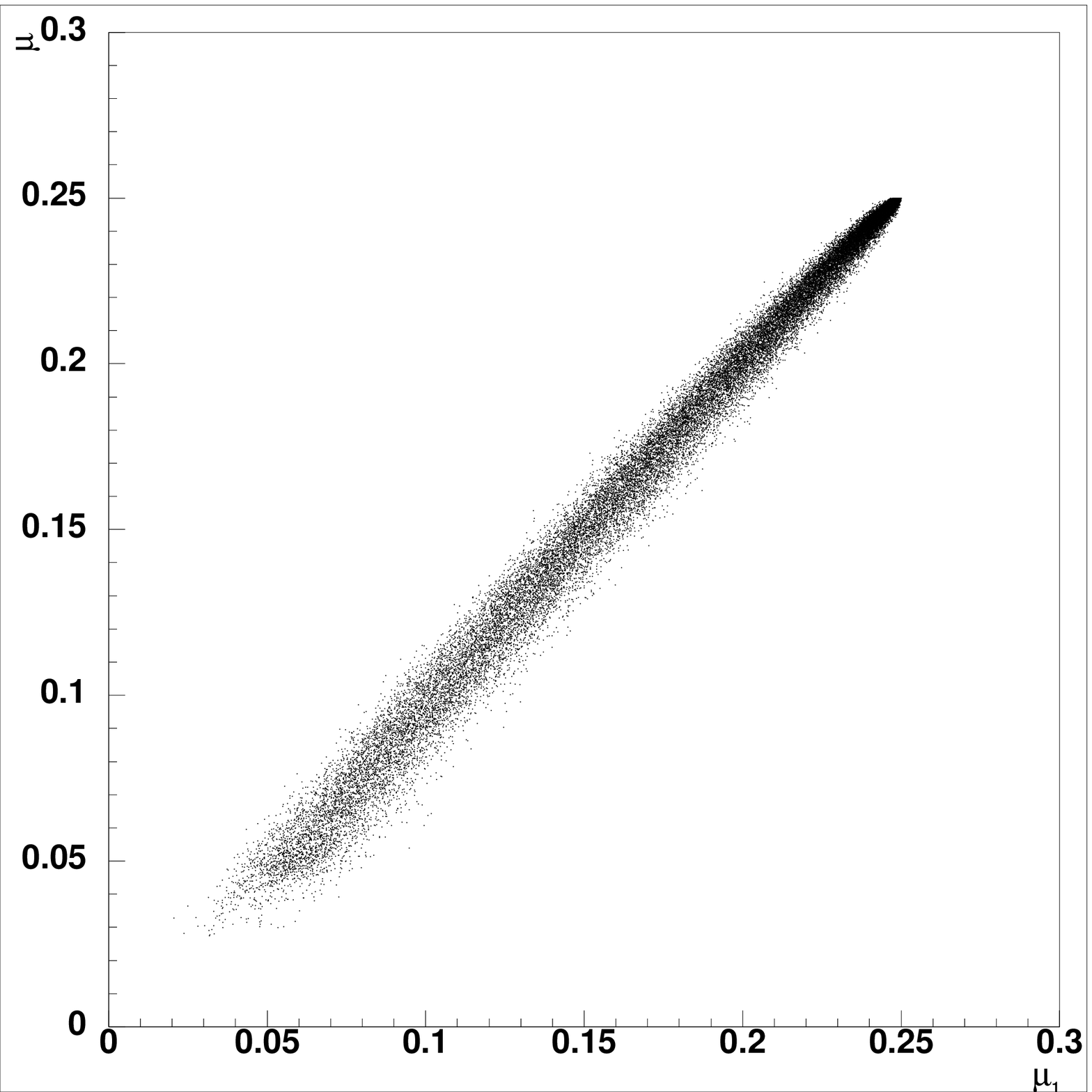,width=75mm}
{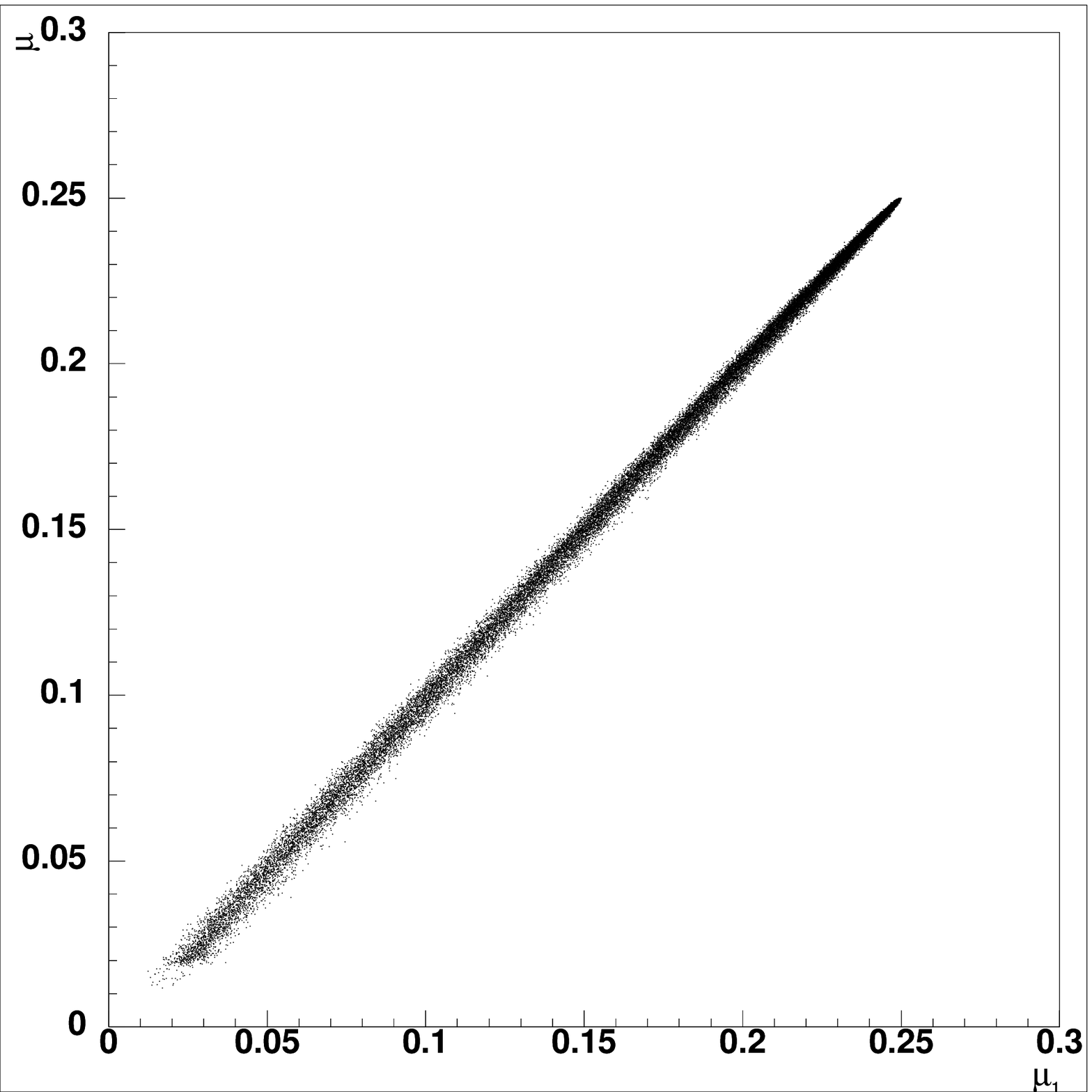,width=75mm}
{$\mu$ (ordinate) vs $\mu_1$ (abscissa) of 3-jet events of 80 GeV c.m. energy.\label{mu80}}
{$\mu$ (ordinate) vs $\mu_1$ (abscissa) of 3-jet events of 200 GeV c.m. energy.\label{mu200}}
\FIGURE{\epsfig{file=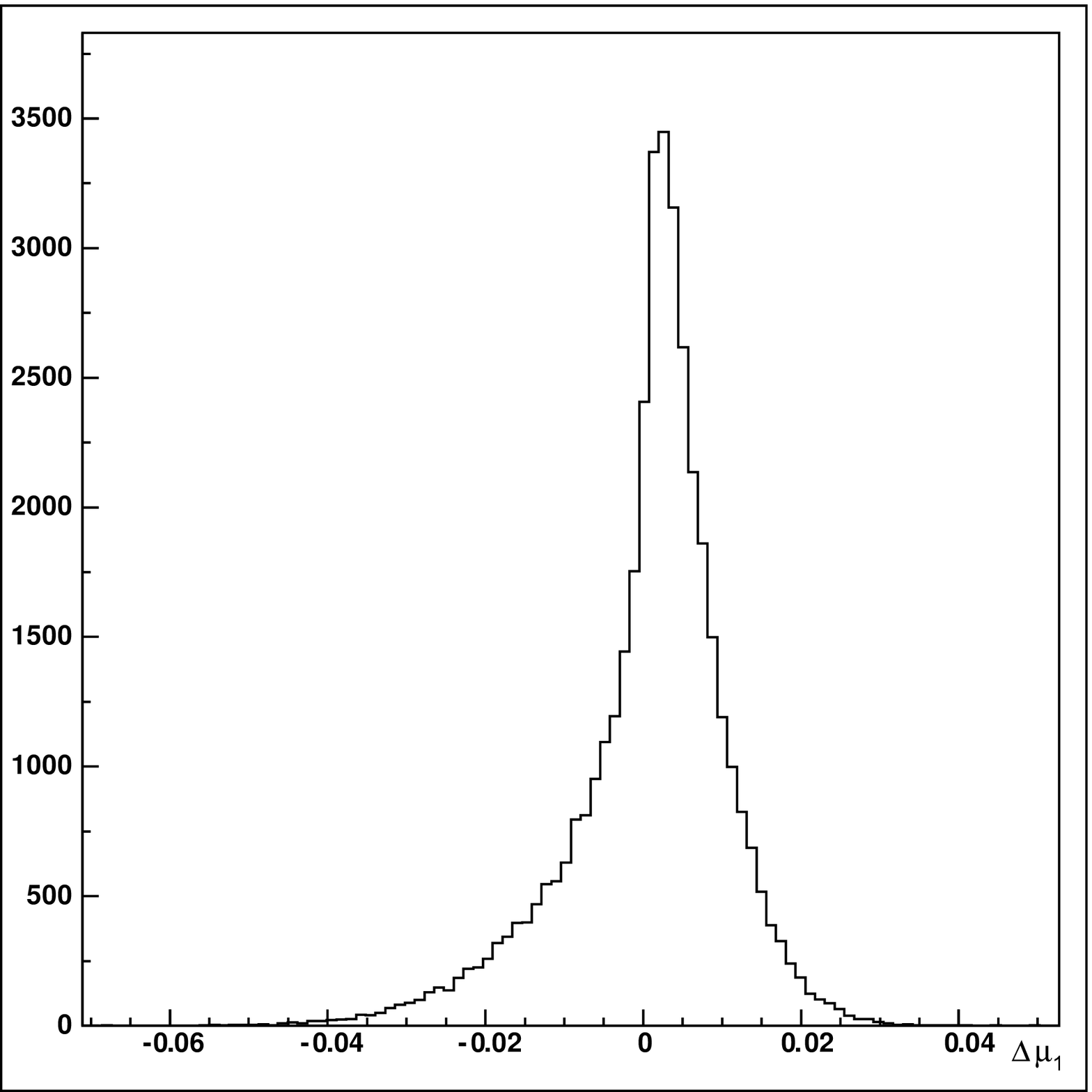,width=110mm}
\caption{Distribution of $\Delta\mu_1 = \mu-\mu_1$ (\ref{4.12}) for 40GeV c.m. energy}
\label{delta}
}
\FIGURE{\epsfig{file=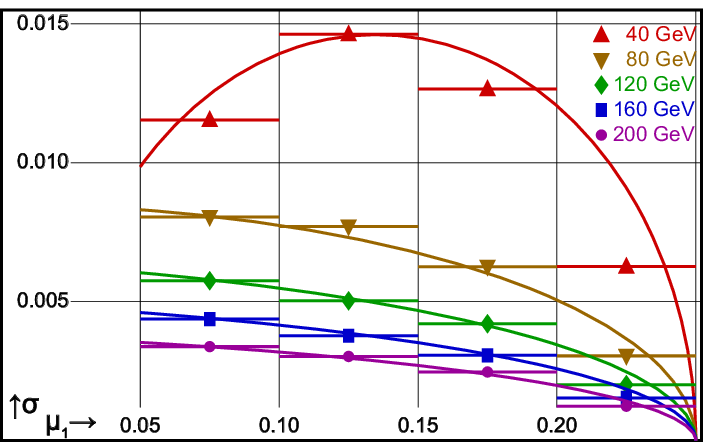,width=110mm}
\caption{Standard deviation $\sigma$ (ordinate) vs $\mu_1$ (\ref{apprmu})  (abscissa) of 3-jet events of different c.m. energies. The curves are fitted ellipses for the parameterization of $\sigma$ (\ref{4.13}).} 
\label{sigmu1}}

\fig{delta} displays the distribution of 
\begin{equation}
\Delta \mu_1  = \mu  - \mu_1  \label{4.12}
\end{equation}
obtained from 40000 generated events at 40 GeV c.m. energy. Its mean value $\left\langle {\Delta \mu } \right\rangle  =  0.000771$ represents the systematic deviation of $\mu_1$ ftom $\mu$. The corresponding standard deviation $\sigma  = 0.01012$ is due to the hadronization process of the jets. 
\FIGURE{\epsfig{file=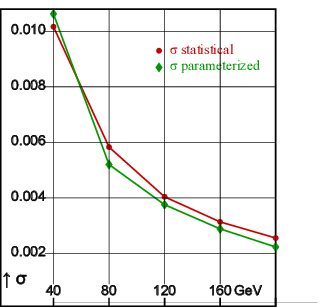,width=65mm}
\caption {The standard deviation $\sigma$ (ordinate) obtained by the statitical distribution, and by the parametrized expression, vs the c.m. energy (abscissa)}. \label{sigepar}}

In order to be able to use the standard deviation, one needs to express it by the measured quantities: $\mu_1$~(\ref{apprmu}) and the c.m. energy $E$. For this purpose, the events of each one of the investigated energies  (40, 80, 120, 160, 200 GeV) were divided into  four regions of $\mu_1$: 0.05 to 0.10; 0.10 to 0.15; 0.15 to 0.20; 0.20 to 0.25. The region below 0.05 was not used, because of the depletion of events there. For each energy and region of $\mu_{1}$, the appropriated standard deviation was calculated. 

The standard deviation $\sigma$ as function of the energy and $\mu_1$, obtained that way, is displayed in \fig{sigmu1}. The solid lines represent ellipses, fitted by "Minuit" \cite{Minuit}, yielding $\chi^{2}=3.1$ for 15 degrees of freedom, and are expressed by:
\begin{equation}
\sigma  = \beta \sqrt {\xi \left( {2a - \xi } \right)} \left\{ \matrix{
  \xi  = .25 - \mu _1  \hfill \cr 
  \beta  = {c \over {E + \varepsilon }} \hfill \cr 
  a = a_0 \left[ {E\left( {E + a_1 } \right) + a_2 } \right] \hfill \cr}  \right\}
\label{4.13}
\end{equation}
where the c.m. energy of the event \emph{E} is in GeV, and the numerical values of the parameters are in the appropriate units:
$c = 1.812 \pm .347$; $\varepsilon  =  - 25.74 \pm 2.85$; $a_0  =  - .00001485 \pm .00000486$; $a_1  =  - 355.3 \pm 27.4$; $a_2  = 4864 \pm 2839$ . Ellipses were chosen just for practical reasons, in order to obtain a simple formula for interpolating the 
$\mu_1$ and $E$ values. The fixed point of $\sigma = 0$ at $\mu_1 = 0.25$ for all energies is backed up by the trend of the data points, consistent with being a stationary point corresponding to the maximal $\mu_1$. The obtained $\sigma$ values from (\ref{4.13}) are due to the hadronization process alone, and the experimentalist should take care of adding the errors of measurement. $E=$ 40 GeV is used as the lower limit for the present analysis and any extrapolation below this energy could be erroneous. This limit was chosen, since for lower energies, the standard deviation increases sharply, and the use of $\mu_1$ becomes non-practical. On the other hand an  extrapolation to higher than 200 GeV energies is expected to be safe. $\mu _1<$ 0.05 was not used for the fits, and it is advisable to make at least such cut in the data, in order to avoid two-jet events.

A test of the  standard deviation function (\ref{4.13}) was done by calculating $\sigma$ for $N=40000$ generated events at each given energy, by use of
\begin{equation}
\sigma  = \sqrt {{{\sum\limits_{n = 1}^N {\sigma _n^2 } } \over N}} 
\label{4.17}
\end{equation}
where $\sigma_n$  corresponds to the value of $\sigma$ (\ref{4.13}) for the $n^{th}$ event, rather than from the deviation of the statistical distribution, presented previously at \fig{sige1}. The results, superimposed on the statistical calculation, are shown at \fig{sigepar} for comparison. Both exhibit a pattern close to each other, with deviations of less than 10\%. This is satisfactory in view of the two different procedures, giving us confidence in the parameterization (\ref{4.13}).  

\subsection{Tensor of rank three}

It was shown that the use of the rank-3 tensor, for idealized jets, supplies together with the tensor of rank-2, a signature of a 3-jet event. In such a case the  eigenvalues $\nu_i$ of the matrix $R_{ij}$ (\ref{3.05}) fulfill (\ref{3.10}), which is reproduced here for convenience: 
\begin{equation}
\left. \matrix{
  \nu _1  = \nu _2  = 2\lambda _1^2 \lambda _2^2  = 2\mu ^2  \hfill \cr 
  \nu _3  = \lambda _3  = 0 \hfill \cr}  \right\} \label{5.01}
\end{equation}
where $\lambda_i$ are the eigenvalues of the tensor of rank-2 $Q_{ij}$ (\ref{1.01}). 
Real events do not fulfill $\nu_1=\nu_2$ anymore, but they remain close to each other, and $\nu_3$ does not vanish.  
\DOUBLEFIGURE[ht]{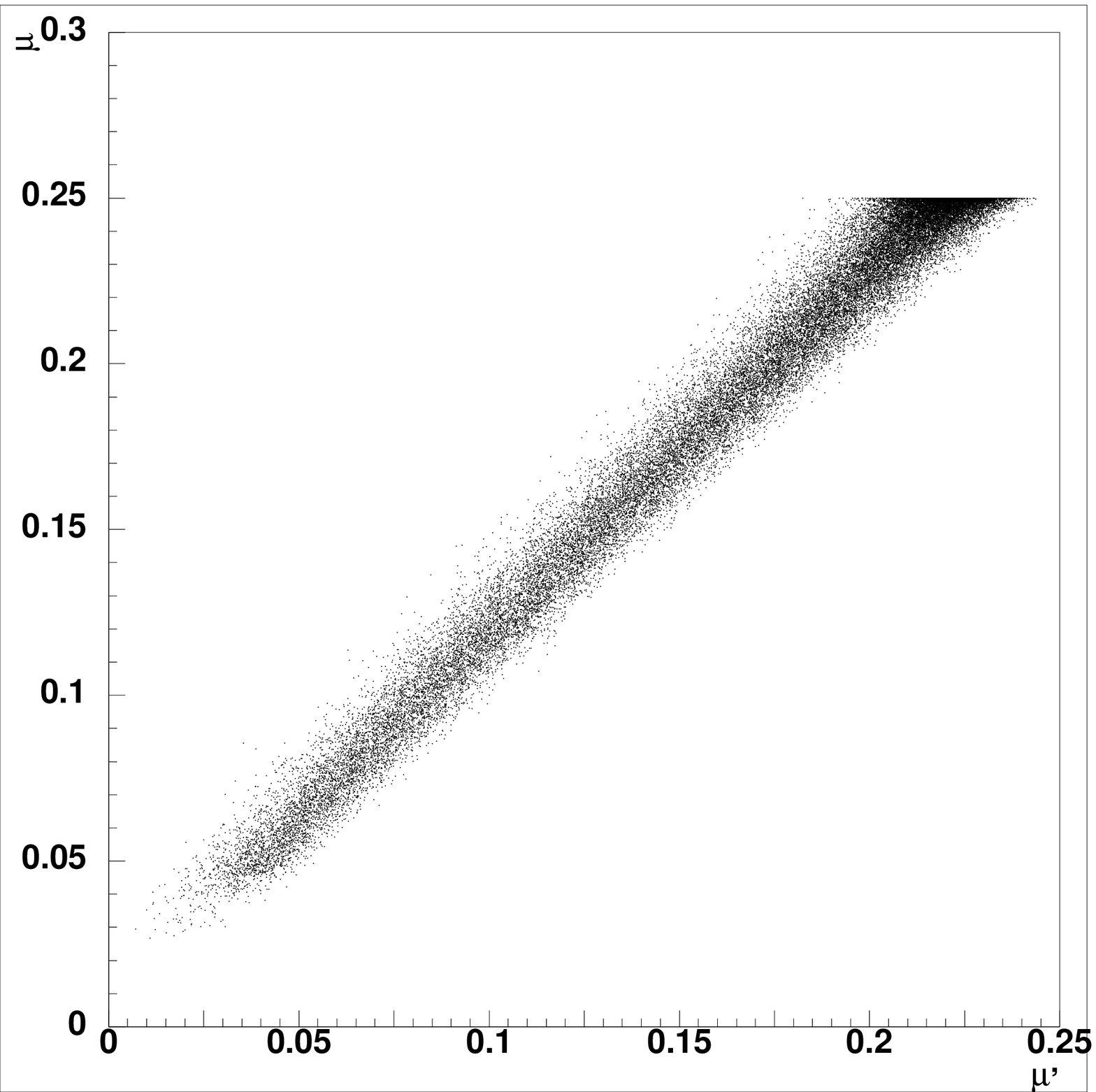,width=75mm}
{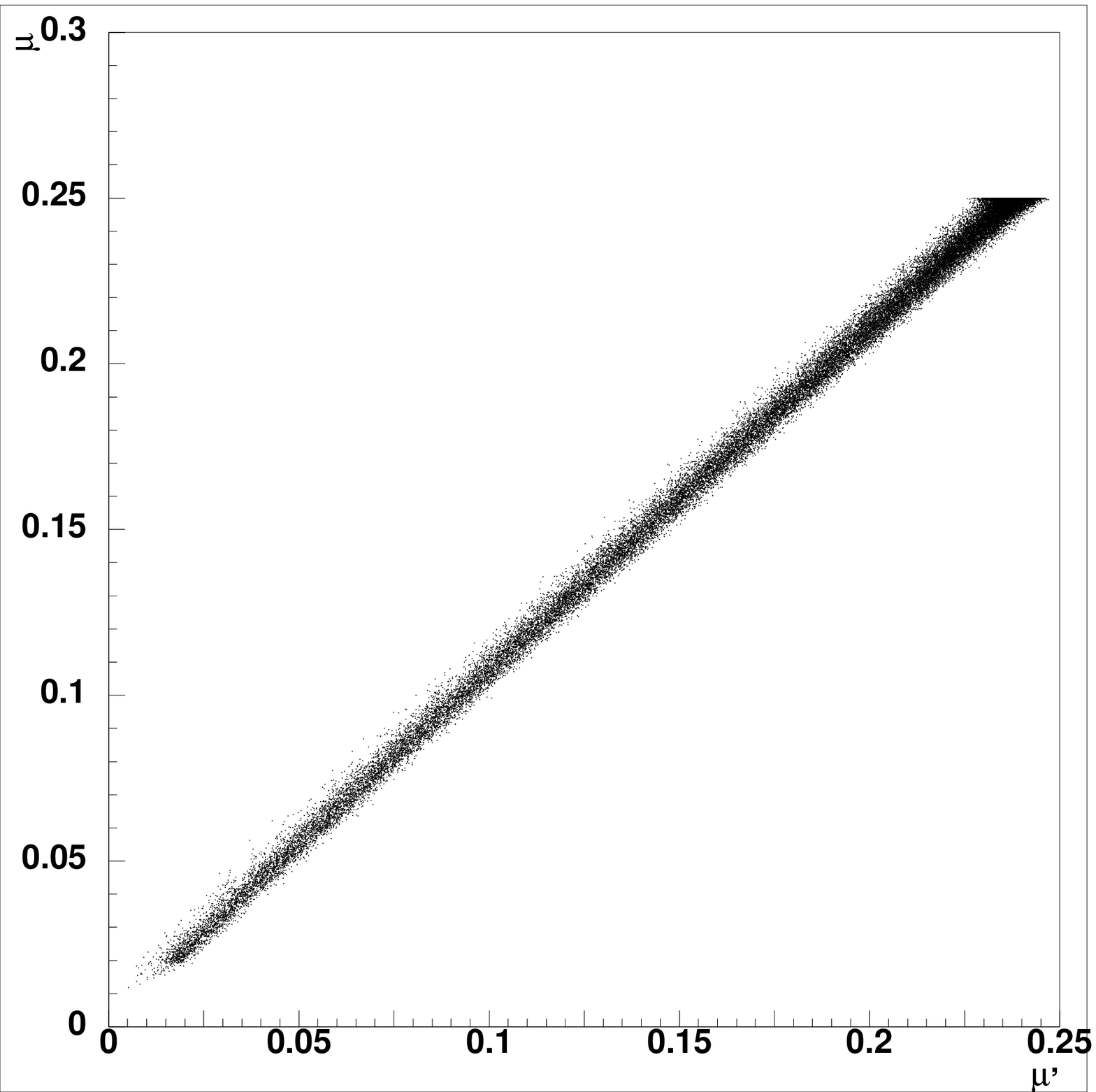,width=75mm}
{$\mu$ (ordinate) vs $\mu'$ (abscissa) of 3-jet events with 80 GeV c.m. energy.\label{mu3t80}}
{$\mu$ (ordinate) vs $\mu'$ (abscissa) of 3-jet events with 200 GeV c.m. energy.\label{mu3t200}}
\FIGURE{\epsfig{file=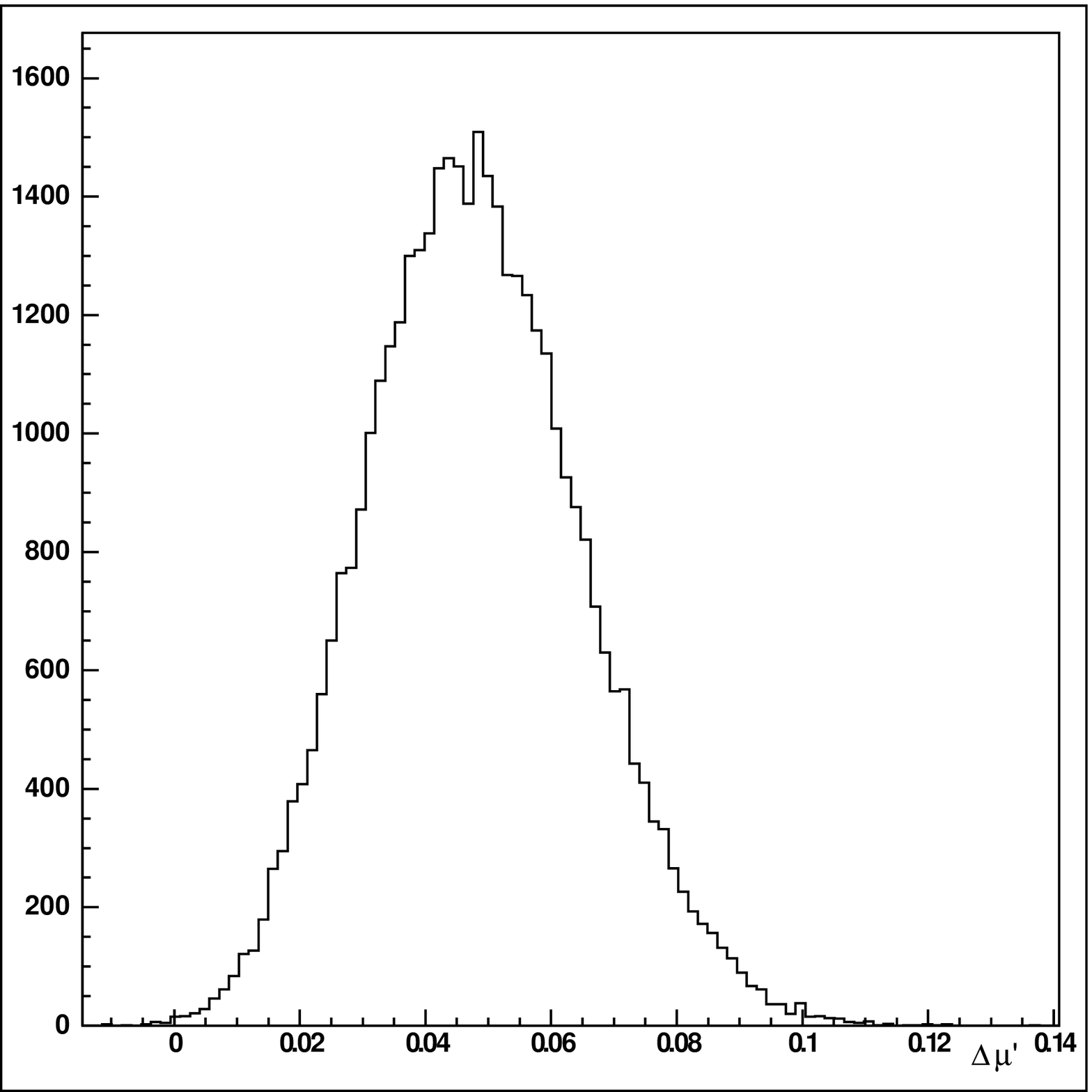,width=100mm}
\caption{Distribution of $\Delta\mu'=\mu-\mu'$ for generated 3-jet events of 40 GeV c.m. energy} 
\label{delta3t40}}

As a first approximation of the value of $\mu$ obtained from the tensor of rank-3, we take according to (\ref{5.01}):
\begin{equation}
\nu _1  \ge \nu _2  \ge \nu _3  \ge 0\quad \quad {\rm{and}}\quad \quad \mu ' = {{\sqrt {\nu _1  + \nu _2 } } \over 2}
\label{5.02}
\end{equation}
The scattergrams in \fig{mu3t80} and \fig{mu3t200}, based on 40000 generated 3-jet events each, display the ideal $\mu$ values (ordinate) versus $\mu'$ (abscissa), for c.m. energies of 80 and 200 GeV. Obviously the vast majority of events have $\mu>\mu'$. This is also seen in \fig{delta3t40}, where the 
\begin{equation}
\Delta\mu'={\mu-\mu'}\label{5.021}
\end{equation}
of 40 GeV events are plotted. In addition, the scattergrams show a depletion of events near $\mu'=0.25$. In \fig{delta3t40} the mean value is $\langle\Delta\mu'\rangle=0.0479$, and the standard deviation is $\sigma=0.0171$.
\FIGURE{\epsfig{file=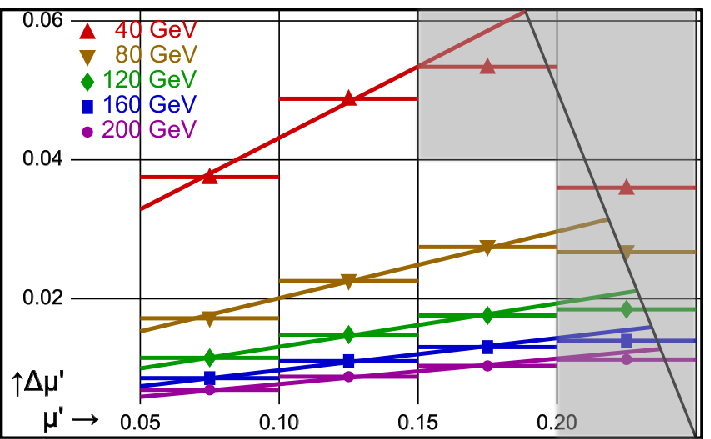,width=110mm}
\caption{$\Delta\mu'$ (\ref{5.021}) (ordinate) vs $\mu'$ (\ref{5.02}) (abscissa) of 3-jet events for different energies.} 
\label{mu3tdel}}
  
In order to obtain a more correct approximation to the value of $\mu$ as a function of the measured quantities $\mu'$ and the c.m. energy $E$, we parametrized $\Delta\mu'$ (\ref{5.021}) as a function of these variables. \fig{mu3tdel} shows the average values of $\Delta\mu'$ for given spans of $\mu'$, and for different energies. In addition to the span of $\mu'$ below the value 0.05, the shaded parts are also not used, in order to avoid the $\mu'$ values from the depleted regions.
The straight lines, corresponding each to a given energy, were obtained from fitting by "Minuit" \cite{Minuit}, yielding $\chi^{2}=0.19$ for 11 degrees of freedom. The results are:
\begin{equation}
\Delta_2  \left( {\mu' ,E} \right) = \mu  - \mu'  = {c \over {E + \varepsilon }}\left( {\mu' + a} \right)
\label{5.03}
\end{equation}
where the energy $E$ is in GeV, and the numerical values of the parameters are: $c = 7.167 \pm 1.204$; $\varepsilon  =  -5.16 \pm 4.09$ and $a  = 0.1098 \pm .0404$. As in the case of the parameterization (\ref{4.13}), the extrapolation to energies lower than 40 GeV is not safe. The more correct approximation of $\mu$ as a function of the measured values is therefore
\begin{equation}
\mu _{2}  = \mu'  + \Delta _2 \left( {\mu' ,E} \right)
\label{5.04}
\end{equation}
The straight line $\Delta\mu' + \mu' =  0.25$ from \fig{mu3tdel} limits the maximal value of $\mu _{2}$  to the physical limit of $0.25$, meaning that \emph{if one obtains $\mu _{2} > 0.25$, the value of 0.25 should be used}. Otherwise, there are no restrictions on the $\mu'$ values, for using (\ref{5.04}).
\DOUBLEFIGURE[ht]{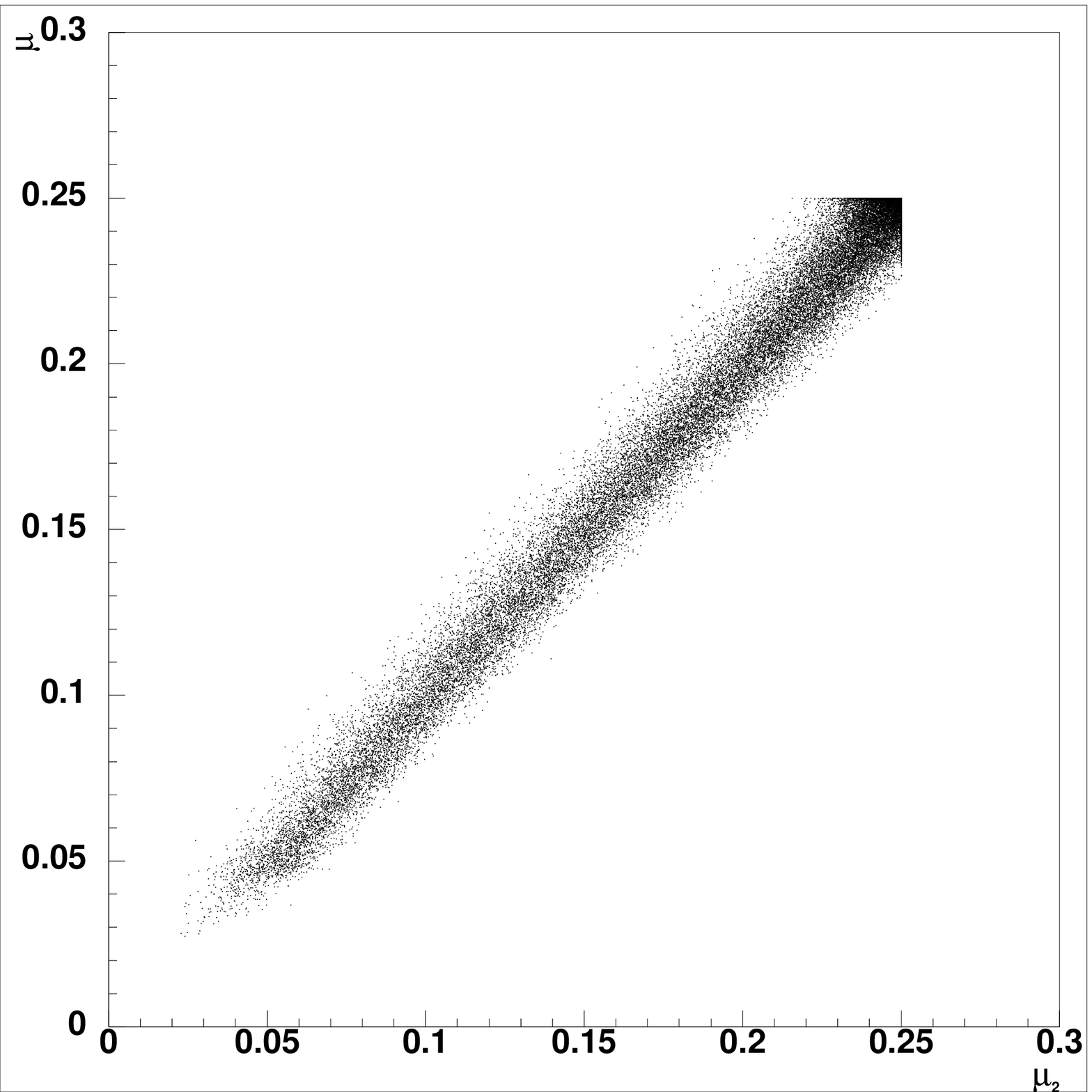,width=75mm}
{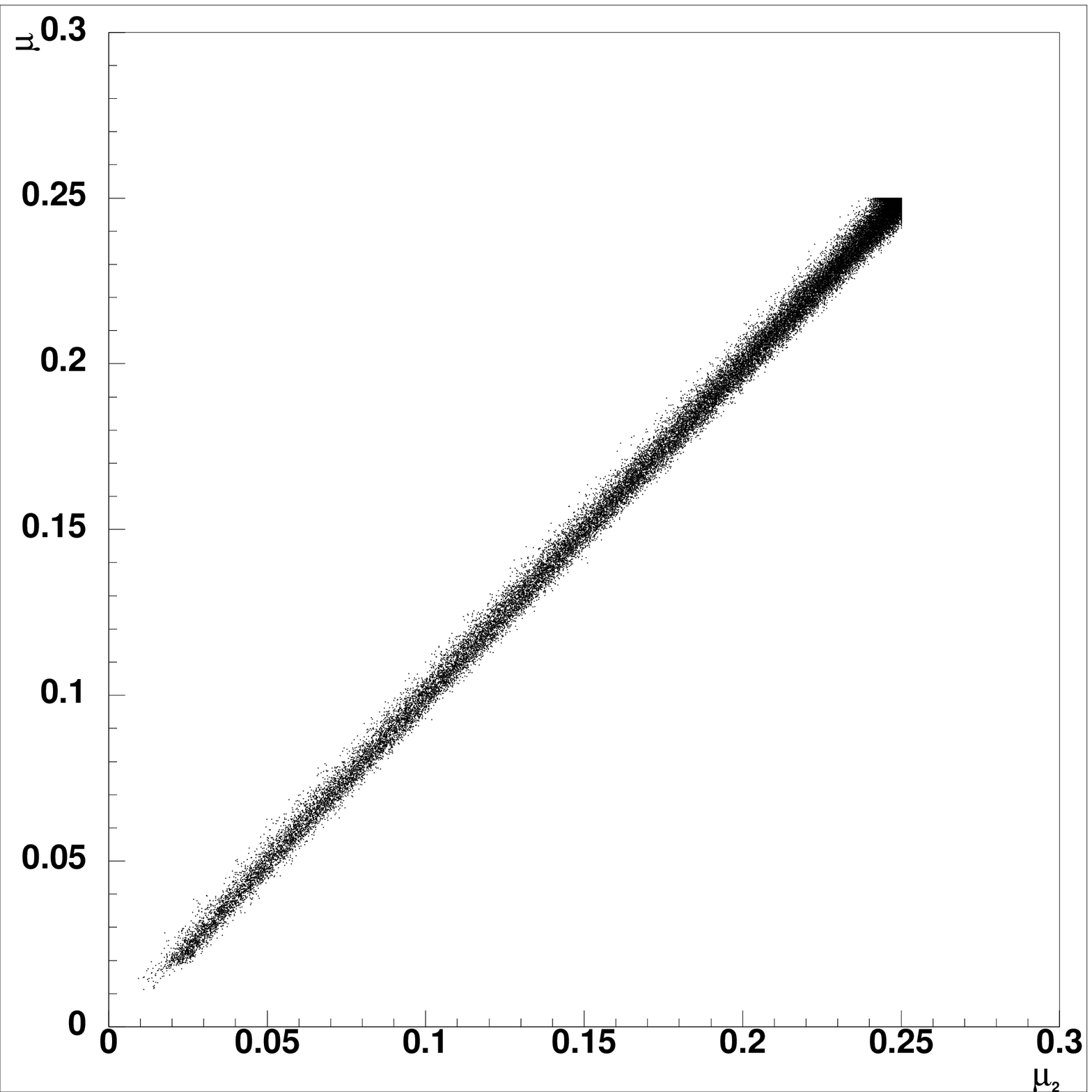,width=75mm}
{$\mu$ (ordinate) vs $\mu_{2}$ (abscissa) of 3-jet events with 80 GeV c.m. energy.\label{cormu3t80}}
{$\mu$ (ordinate) vs $\mu_{2}$ (abscissa) of 3-jet events with 200 GeV c.m. energy.\label{cormu3t200}}

After applying the correction of (\ref{5.04}) one obtains the scatter plots \fig{cormu3t80} and \fig{cormu3t200}, which should be compared with \fig{mu3t80} and \fig{mu3t200}, using uncorrected $\mu'$ values. 

The display of the corrected value
\begin{equation}
\Delta \mu _2  = \mu  - \mu _2 
\label{5.041}
\end{equation}
presented in \fig{cordelta3t40} should be compared to the - uncorrected of \fig{delta3t40}. After correcting, one obtains improved values of the mean: $-0.00248$, and of the standard deviation: $0.01573$. Even so, these values are worse than those obtained using the tensor of rank-2; for comparison see \fig{dele1} and \fig{sige1}. For this reason the values of $\mu_{1}$ are more reliable than those of $\mu_{2}$. 
\FIGURE{\epsfig{file=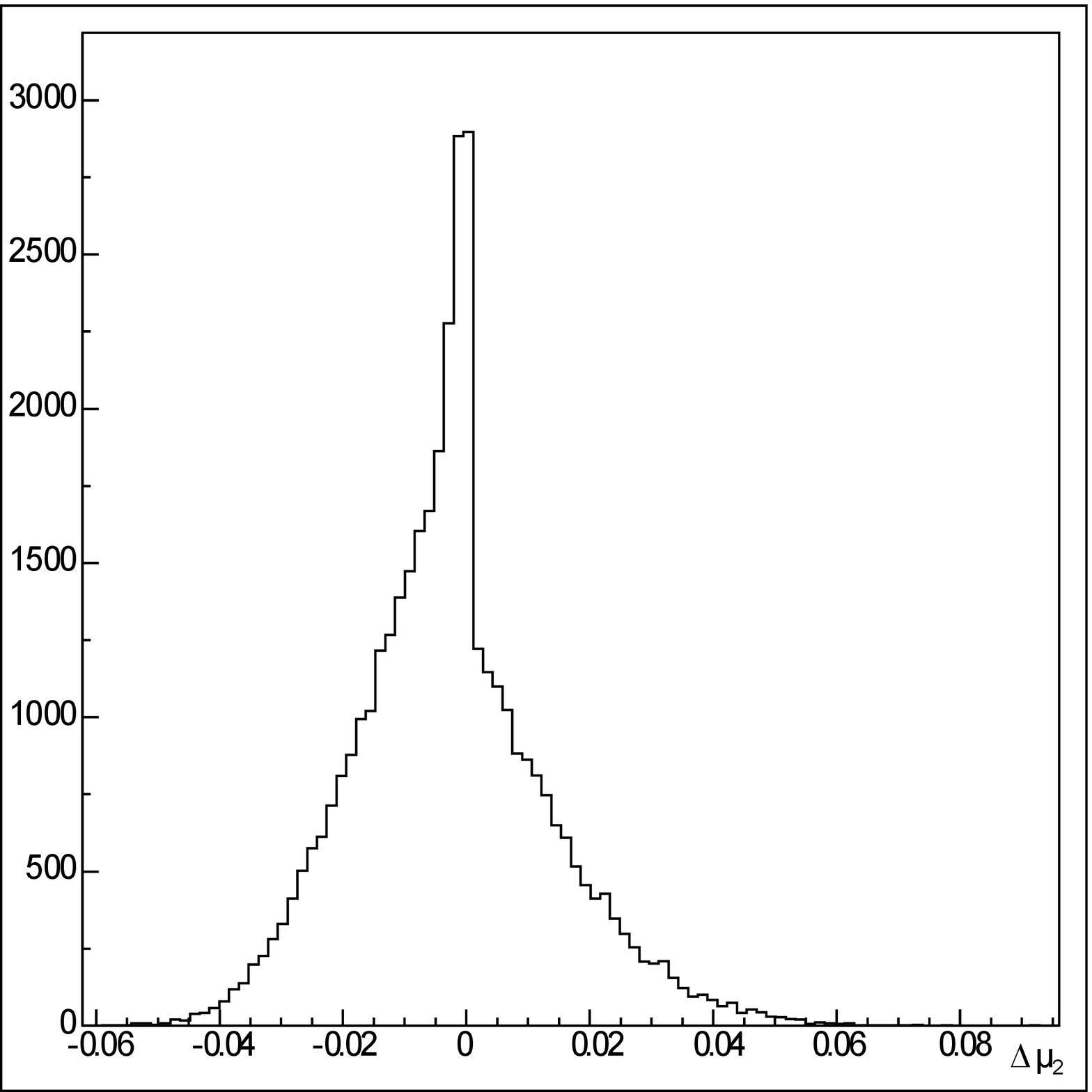,width=95mm}
\caption{Distribution of $\Delta\mu_2=\mu-\mu_{2}$ for generated 3-jet events of 40 GeV c.m. energy} 
\label{cordelta3t40}}
\DOUBLEFIGURE[ht]{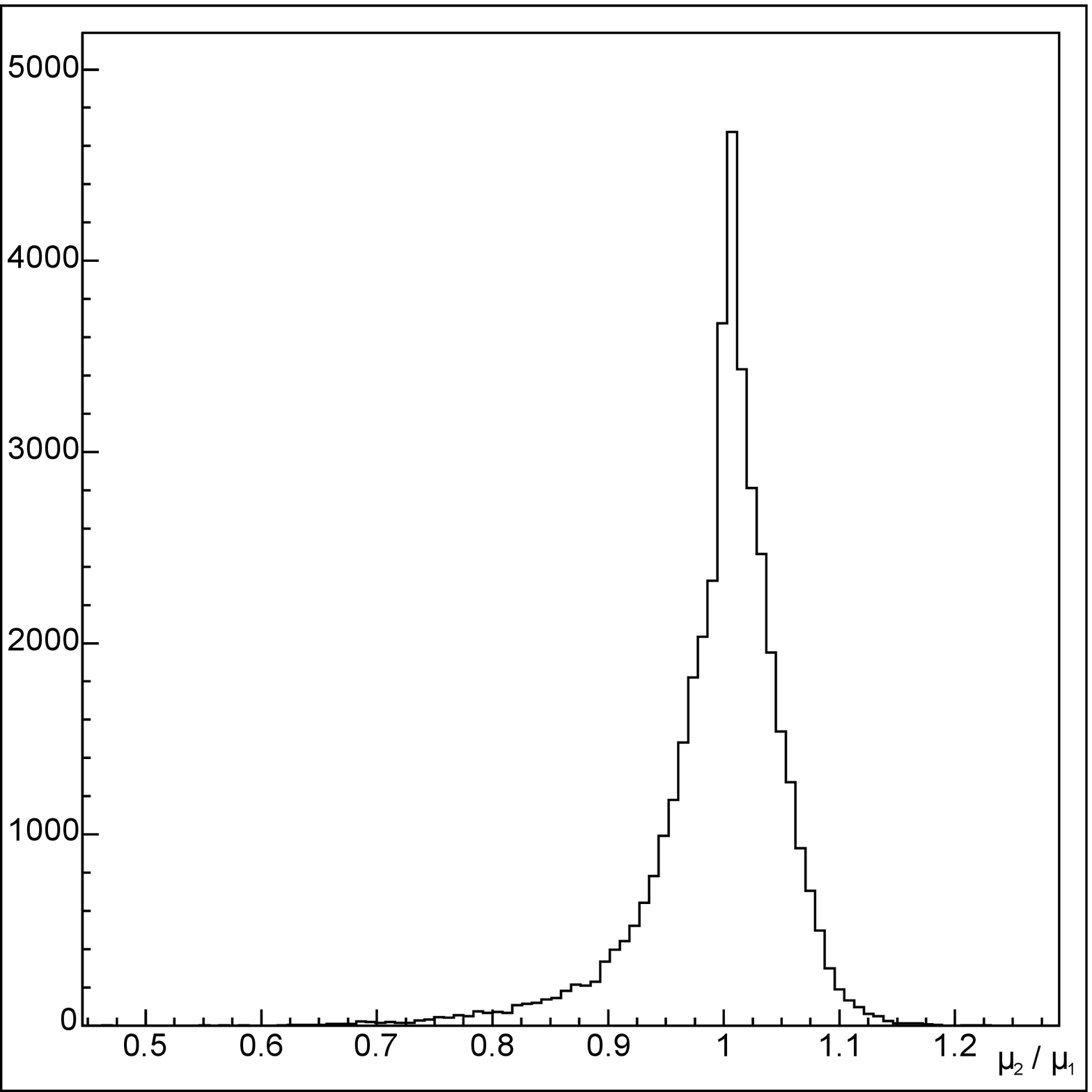,width=86mm}
{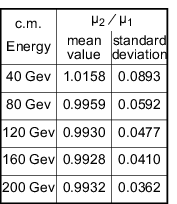,width=45mm}
{Distribution of $\mu_{2}/\mu_{1}$ for generated 3-jet events of 80 GeV c.m. energy.\label{ratio80}}
{Mean value and standard deviation of $\mu_{2}/\mu_{1}$ for different c.m. energies.\label{ratiolist}}
Nevertheless, their ratio is close to unity, as expected from (\ref{5.01}), and shown in \fig{ratio80}, for energy of 80 GeV. The mean value in this case is 0.9959 and the standard deviation is 0.0592, which makes it very convenient for picking up 3-jet events over a large background. This kind of behaviour is characteristic for all the studied energies, from 40 to 200 GeV, as shown in \fig{ratiolist}.

\subsection{Real data}
This method differs from the generally accepted ones for jet analysis, therefore it is necessary to discuss how to implement it on real data. The method is applicable for 3-jet events only, therefore we start by reviewing the selection criteria for such events.

The momentum tensors require to have all the particles originating from the event's vertex, boosted to their common centre of mass system. An event should be accepted only if the centre of mass energy $E$ is above a limit of at least 40GeV, in order to obtain accurate enough results. The momentum tensor of rank-2 $Q_{ij}$ (\ref{1.01}) should be calculated and its eigenvalues ($\lambda$'s) obtained and ordered  (\ref{1.05}). The value of $\mu_1$ (\ref{apprmu}), should be calculated and used as an approximation to the idealized $\mu$ value (\ref{2.01}). By definition its value is between 0 and 0.25 . An event should be accepted only if its value is above a limit of at least 0.05, in order to avoid 2-jet events. 

The momentum tensor of rank three $Q_{ijk}$ (\ref{3.01}) should be calculated, and its corresponding tensor of rank two $R_{ij}$~(\ref{3.05}-\ref{3.06}) obtained. The eigenvalues ($\nu$) of the latter should be ordered, and the value of $\mu'$ (\ref{5.02}) calculated. This value is influenced by the hadronization, and needs to be transformed to $\mu_2$ (\ref{5.03}-\ref{5.04}), following the related instructions in the text. As a result the distribution of ($\mu_1-\mu_2$) should include a peak centred close to the zero, and a background mostly toward the positive values. Also the ($\mu_2/\mu_1$) distribution should show a peak pointing to a value close to  one with a background, mostly to the left. These peaks correspond to the 3-jet events, and the second one is expected to be sharper, therefore more suitable for the selection. 

The transformation to $\mu_2$ just described, corresponds to the hadronization of the event generator used in the paper. In order to  obtain more realistic results, the parameters "$a$" and "$c$" from (\ref{5.03}) could be tuned, in order to move the centre of the peaks to the expected values.

After the 3-jet events are selected, the present study makes possible the direct measurement the hadronization corrections of the $\mu$~value (\ref{apprhadcorr}, \ref{hadcorlimit}), which are of theoretical and practical importance. This may improve the choice of a more realistic event generator for Monte-Carlo studies.

Comparisons between the results based on the analysis of 3-jet events using the momentum tensors, and between the identified jets by the clustering methods, can be made, especially on event by event basis. Of course the identified jets obtained from the clustering should be boosted to their centre of mass system first. Their momenta could be used for calculating $\mu$ of the idealized jets (\ref{2.06}), and compared with $\mu_1$ (\ref{apprmu}) from the tensor analysis. The relative efficiency of both methods could also be studied.  

 \section{Summary and conclusions} \label{summary}
The rank-2 momentum tensor, calculated from the momenta of the jets, in a 3-jet event, provides important characteristics of the event. This is not so, if the tensor's calculation is based on all of the particles originating from the vertex of the event, without assigning them to the jets. The paper shows that for high centre of mass energy ($\ge40$  GeV), these characteristics could be retrieved within some calculable error limits. The method of this procedure was constructed and tested by the analysis of Monte-Carlo generated events.

In addition, the momentum tensor of rank-3 is studied, and it is shown that together with the tensor of rank-2, one obtains an unique signature of 3-jet events. This part is also supported by an analysis of generated events. 

Since the use of the rank-3 momentum tensor, together with this of rank-2, supplies signatures of 3-jet events, it is suggestive to study, if tensors of higher ranks do supply signatures of events with higher number of jets.

\acknowledgments{We are pleased to acknowledge \emph{Shmuel Nussinov} for encouragement and fruitful discussions in different stages of this work, and to \emph{Asher Gotsman} and \emph{Odette Benary} for making the English more representable. Thanks also to (in alphabetic order) \emph{Gideon Bella}, \emph{Yan Benhammou}, \emph{Erez Etzion}, \emph{Noam Hod}, \emph{Ronen Ingbir}, \emph{Sergey Kananov} and \emph{Eugene Levin} for supplying us important technical support. The online image converter \href{http://image.online-convert.com/}{http://image.online-convert.com/} was very handy and efficient. Thanks!}

\appendix
\section{Symmetric planar jets, tensor of rank 2}\label{signature2}
From the definition of an idealized, rotationally symmetric planar event of $N\geq3$ jets (\ref{symNjets}) (\fig{5jets}) and the definition of the momentum tensor of rank-2 (\ref{1.01}), one obtains
\begin{equation}
\left. \matrix{
  Q_{11}  = {1 \over N}\sum\limits_{n = 1}^N {\cos ^2 \left( {{{2\pi n} \over N}} \right)}  = {1 \over {2N}}\sum\limits_{n = 1}^N {\left[ {1 + \cos \left( {{{4\pi n} \over N}} \right)} \right]}  \hfill \cr 
  Q_{22}  = {1 \over N}\sum\limits_{n = 1}^N {\sin ^2 \left( {{{2\pi n} \over N}} \right)}  = {1 \over {2N}}\sum\limits_{n = 1}^N {\left[ {1 - \cos \left( {{{4\pi n} \over N}} \right)} \right]}  \hfill \cr 
  Q_{12}  = Q_{21}  = {1 \over N}\sum\limits_{n = 1}^N {\sin \left( {{{2\pi n} \over N}} \right)\cos \left( {{{2\pi n} \over N}} \right)}  = {1 \over {2N}}\sum\limits_{n = 1}^N {\sin \left( {{{4\pi n} \over N}} \right)}  \hfill \cr}  \right\}
\label{Q2sig2}
\end{equation}
but the final sums from (\ref{Q2sig2}) can be expressed in a closed form, (see e.g. \cite{Gradshteyn} 1.341) : 
\begin{equation}
\left. \matrix{
  \sum\limits_{n = 1}^N {\cos \left( {n\varphi } \right)}  = {{\cos \left( {{{N\varphi } \over 2}} \right)\sin \left( {{{N + 1} \over 2}\varphi } \right)} \over {\sin \left( {{\varphi  \over 2}} \right)}}\; - \;1 \hfill \cr 
  \sum\limits_{n = 1}^N {\sin \left( {n\varphi } \right)}  = {{\sin \left( {{{N\varphi } \over 2}} \right)\sin \left( {{{N + 1} \over 2}\varphi } \right)} \over {\sin \left( {{\varphi  \over 2}} \right)}} \hfill \cr}  \right\}
\label{closedForm}
\end{equation}
which yields for any $N\geq3$
\begin{equation}
\left. \matrix{
  \sum\limits_{n = 1}^N {\cos \left( {{{4\pi n} \over N}} \right)}  = {{\cos \left( {2\pi } \right)\sin \left( {{{N + 1} \over N}2\pi } \right)} \over {\sin \left( {{{2\pi } \over N}} \right)}} - 1 = {{\sin \left( {{{2\pi } \over N}} \right)} \over {\sin \left( {{{2\pi } \over N}} \right)}} - 1 = 0 \hfill \cr 
  \sum\limits_{n = 1}^N {\sin \left( {{{4\pi n} \over N}} \right) = } {{\sin \left( {2\pi } \right)\sin \left( {{{N + 1} \over N}2\pi } \right)} \over {\sin \left( {{{2\pi } \over N}} \right)}} = 0 \hfill \cr 
  Q_{11}  = Q_{22}  = {1 \over 2}\quad {\rm{and}}\quad Q_{12}  = Q_{21}  = 0\quad \quad  \Rightarrow \quad \quad \mu  = {1 \over 4} \hfill \cr}  \right\}
\label{Q11Q22}
\end{equation}

\section{Symmetric planar jets, tensor of rank 3}\label{signature3}
From the definition of an idealized, rotationally symmetric planar event of $N\geq3$ jets (\ref{symNjets}) (\fig{5jets}) and the definition of the momentum tensor of rank-3 (\ref{3.01}), one obtains
\begin{equation}
\left. \matrix{
  Q_{111}  = {1 \over N}\sum\limits_{n = 1}^N {\cos ^3 \left( {{{2\pi n} \over N}} \right)}  \hfill \cr 
  Q_{222}  = {1 \over N}\sum\limits_{n = 1}^N {\sin ^3 \left( {{{2\pi n} \over N}} \right)}  \hfill \cr}  \right\}
\label{10.02}
\end{equation}
In the case of $N$=3, by use of~(\ref{10.02}) and~(\ref{3.09}), one obtains
\begin{equation}
\left. \matrix{
  Q_{111} \left( {N = 3} \right)\; = \;{1 \over 4} \hfill \cr 
  Q_{222} \left( {N = 3} \right)\; = \;0 \hfill \cr}  \right\}\quad  \Rightarrow \quad \mu  = 0.25
\label{10.025}
\end{equation}
as it should for a 3-jet event. 

Since
\begin{equation}
\left. \matrix{
  \cos ^3 \varphi  = {1 \over 4}\left[ {3\cos \varphi  + \cos \left( {3\varphi } \right)} \right] \hfill \cr 
  \sin ^3 \varphi  = {1 \over 4}\left[ {3\sin \varphi  - \sin \left( {3\varphi } \right)} \right] \hfill \cr}  \right\}
\label{10.03}
\end{equation}
one obtains for any $N$
\begin{equation}
\left. \matrix{
  Q_{111}  = {1 \over {4N}}\sum\limits_{n = 1}^N {\left[ {3\cos \left( {{{2\pi n} \over N}} \right) + \cos \left( {{{6\pi n} \over N}} \right)} \right]}  \hfill \cr 
  Q_{222}  = {1 \over {4N}}\sum\limits_{n = 1}^N {\left[ {3\sin \left( {{{2\pi n} \over N}} \right) - \sin \left( {{{6\pi n} \over N}} \right)} \right]}  \hfill \cr}  \right\}
\label{10.04}
\end{equation}
which yields using (\ref{closedForm}), together  with the requirement of $N\geq3$ :
\begin{equation}
\left. \matrix{
  Q_{111}  = {1 \over {4N}}\left[ {{{3\cos \pi \sin \left( {\pi  + {\pi  \over N}} \right)} \over {\sin \left( {{\pi  \over N}} \right)}} + {{\cos \left( {3\pi } \right)\sin \left( {3\pi  + {{3\pi } \over N}} \right)} \over {\sin \left( {{{3\pi } \over N}} \right)}} - 4} \right] = {1 \over {4N}}\left[ {{{\sin \left( {{{3\pi } \over N}} \right)} \over {\sin \left( {{{3\pi } \over N}} \right)}} - 1} \right] \hfill \cr 
  Q_{222}  = {1 \over {4N}}\left[ {{{3\sin \pi \sin \left( {\pi  + {\pi  \over N}} \right)} \over {\sin \left( {{\pi  \over N}} \right)}} - {{\sin \left( {3\pi } \right)\sin \left( {3\pi  + {{3\pi } \over N}} \right)} \over {\sin \left( {{{3\pi } \over N}} \right)}}} \right] = {{\sin \left( {3\pi } \right)\sin \left( {{{3\pi } \over N}} \right)} \over {4N\sin \left( {{{3\pi } \over N}} \right)}} \hfill \cr}  \right\}
\label{10.06}
\end{equation}
$Q_{111}$ and $Q_{222}$ vanish for any $N>3$. For $N=3$,  $Q_{111}$ and $Q_{222}$ are undefined in (\ref{10.06}), and they should be taken from (\ref{10.025}).

\end{document}